\documentclass{article}
\usepackage{waspaa23,amssymb,amsmath,url,times}
\usepackage{color}

\usepackage{graphicx}
\usepackage{bm,cite,subfig}
\usepackage{algorithm,algorithmic}

\newcommand{\minimize}{\mathop{\rm minimize}\limits}

\title{Kernel Interpolation of Incident Sound Field in Region Including Scattering Objects}

\name{Shoichi Koyama$^{1}$,
      Masaki Nakada$^{2}$,
      Juliano G. C. Ribeiro$^{2}$,
      and Hiroshi Saruwatari$^{2}$}
\address{$^1$ National Institute of Informatics, 2-1-2 Hitotsubashi, Chiyoda-ku, Tokyo 101-8430, Japan\\              
         $^2$ The University of Tokyo, 7-3-1 Hongo, Bunkyo-ku, Tokyo 113-8656, Japan\\
          koyama.shoichi@ieee.org\\
}

\begin{document}

\ninept
\maketitle

\begin{sloppy}

\begin{abstract}
A method for estimating the incident sound field inside a region containing scattering objects is proposed. The sound field estimation method has various applications, such as spatial audio capturing and spatial active noise control; however, most existing methods do not take into account the presence of scatterers within the target estimation region. Although several techniques exist that employ knowledge or measurements of the properties of the scattering objects, it is usually difficult to obtain them precisely in advance, and their properties may change during the estimation process. Our proposed method is based on the kernel ridge regression of the incident field, with a separation from the scattering field represented by a spherical wave function expansion, thus eliminating the need for prior modeling or measurements of the scatterers. Moreover, we introduce a weighting matrix to induce smoothness of the scattering field in the angular direction, which alleviates the effect of the truncation order of the expansion coefficients on the estimation accuracy. Experimental results indicate that the proposed method achieves a higher level of estimation accuracy than the kernel ridge regression without separation. 
\end{abstract}


\section{Introduction}
\label{sec:intro}

Techniques for estimating and interpolating an acoustic field from multiple microphone observations are essential in the field of acoustic signal processing. By estimating a continuous pressure distribution over a target region or expansion coefficients of the wave functions around a target position from the observed signals, various applications become feasible, e.g., the visualization of acoustic fields~\cite{Maynard:JASA1985,Bertin:CSAbook2015}, interpolation of room impulse responses~\cite{Mignot:IEEE_J_ASLP2013,Ribeiro:IEEE_ACM_J_ASLP2022}, identification of sound sources~\cite{Park:JASA_J_2005,Teutsch:JASA2006}, capturing sound fields for spatial audio~\cite{Poletti:J_AES_2005,Iijima:JASA_J_2021,Koyama:JAES2023}, spatial active noise control (ANC)~\cite{Zhang:IEEE_J_ASLP2018,Maeno:IEEE_J_ASLP2020,Koyama:IEEE_ACM_J_ASLP2021}, among others. 

Current sound field estimation methods are typically based on the expansion of the captured sound field into the wave functions, namely eigenfunctions of the Helmholtz equation, such as plane wave and spherical wave functions~\cite{Laborie:AESconv2003,Poletti:J_AES_2005,Samarasinghe:IEEE_ACM_J_ASLP2014}. However, these methods basically depend on the empirical setting of the truncation order and expansion center because the sound field is decomposed into a finite number of wave functions around a specific expansion center. Sparsity-based extensions are also investigated~\cite{Chardon:JASA2012,Koyama:IEEE_J_JSTSP2019}, but the estimation is basically performed by an iterative process because the inference operator becomes nonlinear. 

The infinite-dimensional analysis of a sound field is proposed in \cite{Ueno:IEEE_SPL2018} and extended to incorporate prior information on source directions in \cite{Ueno:IEEE_J_SP2021}. This method is free from the empirical setting of truncation order and expansion center. When estimating a pressure field with omnidirectional microphones, the method based on the infinite-dimensional analysis corresponds to the kernel ridge regression with the constraint that the solution satisfies the homogeneous Helmholtz equation~\cite{Ueno:IWAENC2018}. Furthermore, the estimation is performed by a linear operation using its closed-form solution. This method has been applied to spatial audio capturing~\cite{Iijima:JASA_J_2021,Koyama:JAES2023} and spatial ANC~\cite{Koyama:IEEE_ACM_J_ASLP2021}. 

The main drawback of the above-described sound field estimation methods is that the presence of scattering objects inside the target region is not taken into consideration. This is because the spherical wave functions and kernel functions used in these methods are derived under the free-field assumption inside the target region, although the presence of scatterers or reverberation outside the target region is allowed. Thus, the estimation accuracy can significantly deteriorate when the target region contains acoustically non-transparent objects. However, in practical applications, it is sometimes necessary to estimate the incident sound field in the region including scattering objects. For example, in the spatial ANC, the pressure distribution of primary noise sources must be estimated inside the target control region from the microphone measurements around the surface of the region. One or more ANC users will present and move within the target region, and they can be scatterers. Several techniques to estimate the incident sound field in the region including scattering objects have been proposed~\cite{Zotkin:ICASSP2017,Ahrens:IEEE_J_ASLP2022}; however, these methods require prior knowledge or measurements of the properties of the scatterers. Obviously, it will not be always possible to obtain them precisely in advance. 

In this paper, we propose a method to estimate the incident sound field in the region including scattering objects without precise knowledge or measurements of their properties. By jointly estimating the coefficients of the incident field represented by a weighted sum of the kernel functions and the scattering field represented by the finite-dimensional spherical wave function expansion, the incident field is estimated based on the kernel ridge regression with a separation from the scattering field. 
The proposed estimation can still be performed by a linear operation. This means that the estimation can be implemented by a convolution of a finite impulse response (FIR) filter, which is suitable for many practical applications. We also introduce a weighting factor for the expansion coefficients of the scattering field derived on the basis of its smoothness to alleviate the effect of the truncation of expansion order. We conducted numerical simulations in a three-dimensional (3D) space to evaluate our proposed method. 

\vfill

\section{Problem statement and prior work}

\subsection{Problem statement}

Suppose that a region of interest $\Omega \subset \mathbb{R}^3$ is a simply connected open subset of $\mathbb{R}^3$. The sound pressure at the position $\bm{r}\in\Omega$ and angular frequency $\omega\in\mathbb{R}$ is denoted as $u(\bm{r},\omega)$. As shown in Fig.~\ref{fig:sf_sct}, one or more scattering objects of arbitrary shape exist inside a spherical region $\Omega_{\mathrm{sct}} \subset \Omega$. $M$ omnidirectional microphones are arbitrarily placed over $\Omega\backslash\Omega_{\mathrm{sct}}$, whose positions are denoted as $\{\bm{r}_m\}_{m=1}^M$. We denote the $m$th microphone measurement as $s_m$ that is equivalent to $u(\bm{r}_m,\omega)$ plus sensor noise. The pressure field $u$ is represented by a sum of incident and scattering fields, $u_{\mathrm{inc}}$ and $u_{\mathrm{sct}}$, as
\begin{align}
 u(\bm{r},\omega) = u_{\mathrm{inc}}(\bm{r},\omega) + u_{\mathrm{sct}}(\bm{r},\omega).
\label{eq:sf_inc_sct}
\end{align}
Our objective is to estimate the incident field $u_{\mathrm{inc}}$ from the microphone measurements $\{s_m\}_{m=1}^{M}$. Hereafter, $\omega$ is omitted for notational simplicity. 

\begin{figure}[t]
  \centering
  \centerline{\includegraphics[width=0.8\columnwidth]{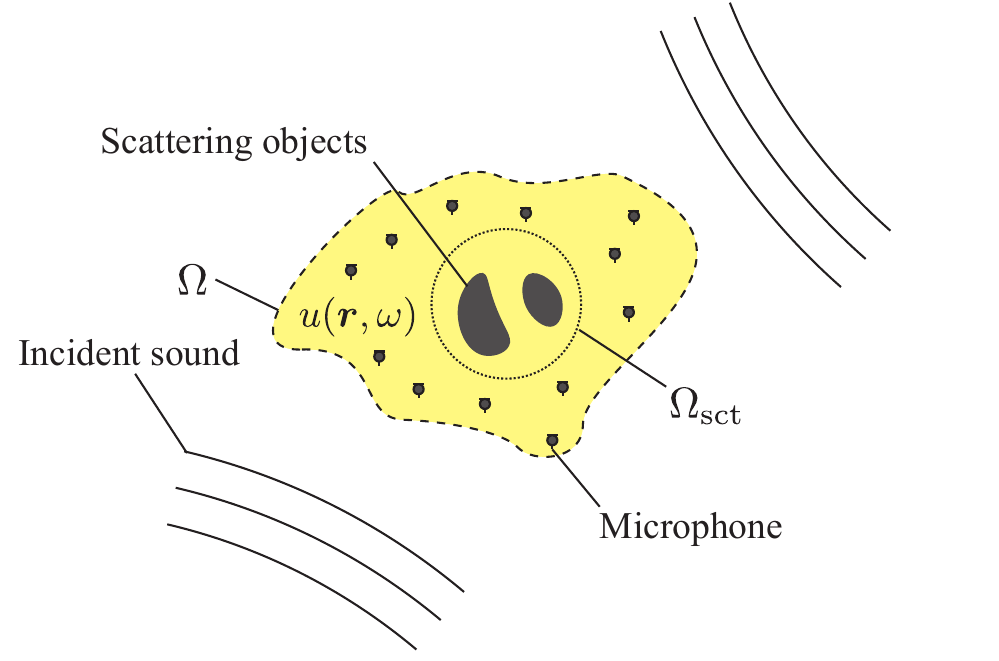}}
  \caption{Estimation of the incident sound field in the region including scattering objects.}
  \label{fig:sf_sct}
\end{figure}

\subsection{Current sound field estimation methods}
\label{sec:priorwork}

When no scattering object exists inside $\Omega$, the incident field $u_{\mathrm{inc}}$, which is equivalent to $u$, can be estimated based on spherical wave function expansion:
\begin{align}
 u_{\mathrm{inc}}(\bm{r}) = \sum_{\nu,\mu} \mathring{u}_{\mathrm{inc},\nu,\mu}(\bm{r}_{\mathrm{o}}) \varphi_{\mathrm{inc},\nu,\mu}(\bm{r}-\bm{r}_{\mathrm{o}}),
\label{eq:sph_exp_int}
\end{align} 
where $\mathring{u}_{\mathrm{inc},\nu,\mu}$ is the expansion coefficients for order $\nu$ and degree $\mu$, $\bm{r}_{\mathrm{o}}\in\Omega$ is the expansion center, and $\varphi_{\mathrm{inc},\nu,\mu}$ is the spherical wave function for interior field defined as
\begin{align}
 \varphi_{\mathrm{inc},\nu,\mu}(\bm{r}) := \sqrt{4\pi} j_{\nu}(k\|\bm{r}\|) Y_{\nu,\mu}\left( \bm{r} / \|\bm{r}\| \right) 
\end{align}
with the $\nu$th-order spherical Bessel function $j_{\nu}(\cdot)$, wave number $k$ ($:=\omega/c$ with sound velocity $c$), and spherical harmonic function $Y_{\nu,\mu}(\cdot)$. The factor $\sqrt{4\pi}$ is multiplied so that $\mathring{u}_{\mathrm{inc},0,0}(\bm{r}_{\mathrm{o}})$ corresponds to $u(\bm{r}_{\mathrm{o}})$. Here, the summation for $\nu$ and $\mu$ represents $\sum_{\nu,\mu} := \sum_{\nu=0}^{\infty} \sum_{\mu=-\nu}^{\nu}$. The expansion coefficients up to a predefined truncation order $N$ can be estimated from the microphone measurements by solving a linear equation constructed by $\{s_m\}_{m}$, $\{\varphi_{\mathrm{inc},\nu,\mu}(\bm{r}_m-\bm{r}_{\mathrm{o}})\}_{m,\nu,\mu}$, and $\{\mathring{u}_{\mathrm{inc},\nu,\mu}\}_{\nu,\mu}$~\cite{Laborie:AESconv2003,Poletti:J_AES_2005}. Then, the pressure field $u_{\mathrm{inc}}$ inside $\Omega$ can be reconstructed by using the estimated expansion coefficients based on \eqref{eq:sph_exp_int}. Note that the empirical setting of the truncation order $N$ and expansion center $\bm{r}_{\mathrm{o}}$ is necessary for this estimation procedure. 

To avoid the empirical setting of truncation order and expansion center, the kernel ridge regression for a sound field can be applied~\cite{Ueno:IEEE_SPL2018,Ueno:IEEE_J_SP2021}, which is a special case of the analysis based on infinite-dimensional spherical wave function expansion~\cite{Ueno:IWAENC2018}. The kernel function is formulated so that the function space to seek a solution is constrained to the solution space of the homogeneous Helmholtz equation. Based on the representer theorem~\cite{Scholkopf:COLT2001}, the pressure distribution $u$ can be represented as a weighted sum of the reproducing kernel functions as
\begin{align}
 u(\bm{r}) &= \sum_{m=1}^M \alpha_m \kappa(\bm{r},\bm{r}_m), 
\label{eq:rep_thrm}
\end{align}
where $\{\alpha_m\}_{m=1}^M$ is the weights, and $\kappa$ is the kernel function. The kernel function is defined as
\begin{align}
 \kappa(\bm{r}_1, \bm{r}_2) = j_0 \left( \left[ \left( \mathrm{j}\rho\bm{\eta}_{\mathrm{pr}} - k(\bm{r}_1-\bm{r}_2) \right)^{\mathsf{T}} \left( \mathrm{j}\rho\bm{\eta}_{\mathrm{pr}} - k(\bm{r}_1-\bm{r}_2) \right) \right]^{\frac{1}{2}}\right),
\label{eq:kernel_dir}
\end{align}
where $\bm{\eta}_{\mathrm{pr}}\in\mathbb{S}_2$ is the prior information on the source direction, and $\rho$ is the weighting parameter for the prior information. When no prior information on the source direction is available, $\rho$ is set to $0$, and the kernel function is simplified as
\begin{align}
 \kappa(\bm{r}_1, \bm{r}_2) = j_0 \left( k \| \bm{r}_1-\bm{r}_2 \| \right).
\label{eq:kernel_diffuse}
\end{align}
By defining $\bm{\alpha}=[\alpha_1, \ldots, \alpha_M]^{\mathsf{T}}$, $\bm{s}=[s_1,\ldots,s_M]^{\mathsf{T}}$, and 
\begin{align}
 \bm{K} = 
\begin{bmatrix}
 \kappa(\bm{r}_1,\bm{r}_1) & \cdots & \kappa(\bm{r}_1,\bm{r}_M) \\
 \vdots & \ddots & \vdots \\
 \kappa(\bm{r}_M,\bm{r}_1) & \cdots & \kappa(\bm{r}_M,\bm{r}_M) 
\end{bmatrix},
\end{align}
$\bm{\alpha}$ is obtained in a closed form as
\begin{align}
 \bm{\alpha} = (\bm{K} + \lambda \bm{I})^{-1} \bm{s},
\label{eq:krr}
\end{align}
where $\lambda$ is the regularization parameter and $\bm{I}$ is the identity matrix. By using the kernel function defined in \eqref{eq:kernel_dir}, \eqref{eq:kernel_diffuse}, or their weighted sum~\cite{Horiuchi:WASPAA2021}, the estimate obtained by using $\bm{\alpha}$ is constrained to the solution of the homogeneous Helmholtz equation. The kernel ridge regression \eqref{eq:krr} is equivalent to Gaussian process regression because the kernel function has no hyperparameters to learn~\cite{Rasmussen:GPforML}.  

These methods are applicable only when the target region does not contain any scattering objects because the homogeneous Helmholtz equation is assumed to be satisfied inside $\Omega$. One of the simple techniques to alleviate the scattering effects is to employ directional microphones whose minimum-gain direction is directed to the scattering objects~\cite{Ueno:IEEE_SPL2018}; however, it is difficult to develop directional microphones having ideal nulls, especially at low frequencies. Several techniques have been proposed to cancel the scattering effects by measuring or modeling them in advance~\cite{Zotkin:ICASSP2017,Ahrens:IEEE_J_ASLP2022}. However, it is not always possible to precisely measure or model the scattering effects in practical situations. 

\section{Proposed method}

We consider extracting and estimating the incident field $u_{\mathrm{inc}}$ from the measurements $\bm{s}$ in $\Omega$ containing unknown scattering objects, without precise measurement or modeling of their properties in advance. Our approach is based on the representation of the scattering field $u_{\mathrm{sct}}$ in $\Omega\backslash\Omega_{\mathrm{sct}}$ by using spherical wave function expansion. Then, the weights of the kernel functions and the expansion coefficients of the spherical wave functions are jointly estimated. Thus, the incident field $u_{\mathrm{inc}}$ can still be estimated as a closed-form solution. 

\subsection{Model}

The sound field $u$ is represented as the sum of $u_{\mathrm{inc}}$ and $u_{\mathrm{sct}}$ as in \eqref{eq:sf_inc_sct}, and $u_{\mathrm{inc}}$ is represented as a weighted sum of the kernel functions \eqref{eq:rep_thrm}. We represent the scattering field $u_{\mathrm{sct}}$ by a finite-dimensional spherical wave function expansion for the exterior field as
\begin{align}
 u(\bm{r}) &= \sum_{m=1}^M \alpha_m \kappa(\bm{r},\bm{r}_m) + \sum_{\nu,\mu} \mathring{u}_{\mathrm{sct},\nu,\mu}(\bm{r}_{\mathrm{o}}) \varphi_{\mathrm{sct},\nu,\mu} (\bm{r}-\bm{r}_{\mathrm{o}}) \notag\\
&\approx \sum_{m=1}^M \alpha_m \kappa(\bm{r},\bm{r}_m) + \sum_{\nu,\mu}^N \mathring{u}_{\mathrm{sct},\nu,\mu} \varphi_{\mathrm{sct},\nu,\mu} (\bm{r}),
\end{align}
where $\sum_{\nu,\mu}^N:=\sum_{\nu=0}^{N}\sum_{\mu=-\nu}^{\nu}$, $\mathring{u}_{\mathrm{sct},\nu,\mu}$ is the expansion coefficients, and $\varphi_{\mathrm{sct},\nu,\mu}$ is the spherical wave function for exterior field defined as
\begin{align}
 \varphi_{\mathrm{sct},\nu,\mu}(\bm{r}) := \sqrt{4\pi} h_{\nu}(k\|\bm{r}\|) Y_{\nu,\mu}(\bm{r}/\|\bm{r}\|)
\end{align}
with the $\nu$th-order spherical Hankel function of the second kind $h_{\nu}$. Thus, the microphone measurements $\bm{s}$ can be described as
\begin{align}
 \bm{s} = \bm{K\alpha} + \bm{\Phi}_{\mathrm{sct}} \mathring{\bm{u}}_{\mathrm{sct}} + \bm{\varepsilon},
\end{align}
where $\mathring{\bm{u}}_{\mathrm{sct}}\in\mathbb{C}^{(N+1)^2}$ is the vector of $\{\mathring{u}_{\mathrm{sct},\nu,\mu}\}_{\nu,\mu}$, $\bm{\Phi}_{\mathrm{sct}}\in\mathbb{C}^{M \times (N+1)^2}$ is the matrix of $\{\varphi_{\mathrm{sct},\nu,\mu}(\bm{r}_m)\}_{m,\nu,\mu}$, and $\bm{\varepsilon}\in\mathbb{C}^M$ is the Gaussian sensor noise.

\subsection{Optimization problem and its solution}

To estimate $\bm{\alpha}$ from $\bm{s}$, eliminating $u_{\mathrm{sct}}$, we formulate the following joint optimization problem of $\bm{\alpha}$ and $\mathring{\bm{u}}_{\mathrm{sct}}$:
\begin{align}
& \minimize_{\bm{\alpha},\mathring{\bm{u}}_{\mathrm{sct}}} \mathcal{J}(\bm{\alpha},\mathring{\bm{u}}_{\mathrm{sct}}) \notag\\
&\hspace{20pt}:=\left\| \bm{s} - \bm{K\alpha} - \bm{\Phi}_{\mathrm{sct}} \mathring{\bm{u}}_{\mathrm{sct}} \right\|^2 + \lambda_1 \bm{\alpha}^{\mathsf{H}} \bm{K} \bm{\alpha} + \lambda_2 \mathring{\bm{u}}_{\mathrm{sct}}^{\mathsf{H}} \bm{W} \mathring{\bm{u}}_{\mathrm{sct}},
\label{eq:opt_prb}
\end{align}
where $\lambda_1$ and $\lambda_2$ are the regularization parameters, $\bm{W}\in\mathbb{C}^{(N+1)^2\times(N+1)^2}$ is the weighting matrix for the expansion coefficients $\mathring{\bm{u}}_{\mathrm{sct}}$. A specific definition of $\bm{W}$ is given in Sect.~\ref{sec:weight}. 

The optimization problem \eqref{eq:opt_prb} can be solved in a closed form. By solving $\partial \mathcal{J}/\partial \mathring{\bm{u}}_{\mathrm{sct}}^{\ast}=\bm{0}$ and $\partial \mathcal{J}/\partial \bm{\alpha}^{\ast}=\bm{0}$, one can obtain
\begin{align}
 \hat{\mathring{\bm{u}}}_{\mathrm{sct}} &= \left( \bm{\Phi}_{\mathrm{sct}}^{\mathsf{H}} \bm{\Phi}_{\mathrm{sct}} + \lambda_2 \bm{W} \right)^{-1} \bm{\Phi}_{\mathrm{sct}}^{\mathsf{H}} (\bm{s} - \bm{K}\hat{\bm{\alpha}}) \\
 \hat{\bm{\alpha}} &= \left( \bm{K} + \lambda_1 \bm{I} \right)^{-1} \left( \bm{s} - \bm{\Phi}_{\mathrm{sct}} \hat{\mathring{\bm{u}}}_{\mathrm{sct}} \right).
\end{align}
By solving the above simultaneous equation, the estimates $\hat{\mathring{\bm{u}}}_{\mathrm{sct}}$ and $\hat{\bm{\alpha}}$ are obtained as
\begin{align}
 \hat{\mathring{\bm{u}}}_{\mathrm{sct}} =& \left[ \bm{\Phi}_{\mathrm{sct}}^{\mathsf{H}} (\bm{K}+\lambda_1\bm{I})^{-1}  \bm{\Phi}_{\mathrm{sct}} + \frac{\lambda_2}{\lambda_1} \bm{W} \right]^{-1} \notag\\
& \hspace{80pt} \cdot \bm{\Phi}_{\mathrm{sct}}^{\mathsf{H}} (\bm{K}+\lambda_1\bm{I})^{-1} \bm{s}.
\end{align}
and
\begin{align}
 \hat{\bm{\alpha}} 
= \left( \bm{K} + \lambda_1\bm{I} + \frac{\lambda_1}{\lambda_2} \bm{\Phi}_{\mathrm{sct}}\bm{W}^{-1}\bm{\Phi}_{\mathrm{sct}}^{\mathsf{H}} \right)^{-1} \bm{s},
\end{align}
respectively. Thus, the incident field $u_{\mathrm{inc}}$ is obtained by using $\hat{\bm{\alpha}}$ and \eqref{eq:rep_thrm}. 


\subsection{Weighting matrix for inducing smoothness}
\label{sec:weight}

It is possible to set $\bm{W}$ in \eqref{eq:opt_prb} as $\bm{I}$; however, the estimation accuracy can be highly dependent on the truncation order $N$ although the optimal $N$ depends on the geometry of the scatterers and their reflective properties. To alleviate the dependence on $N$, we define $\bm{W}$ to induce smoothness of the scattering field $u_{\mathrm{sct}}$ in the angular direction. Such weighting factors on the expansion coefficients are also used in the context of the interpolation of head-related transfer functions~\cite{Duraiswami:ICASSP2004}. We here define $\bm{W}$ so that the third term of \eqref{eq:opt_prb} corresponds to the following form: 
\begin{align}
 \sum_{m=1}^M \| \nabla u_{\mathrm{sct}}(\bm{r}_m) \|^2 = \bm{u}_{\mathrm{sct}}^{\mathsf{H}} \left( \frac{\partial \bm{\Phi}_{\mathrm{sct}}^{\mathsf{H}}}{\partial\theta} \frac{\partial \bm{\Phi}_{\mathrm{sct}}}{\partial\theta} + \frac{\partial \bm{\Phi}_{\mathrm{sct}}^{\mathsf{H}}}{\partial\phi} \frac{\partial \bm{\Phi}_{\mathrm{sct}}}{\partial\phi} \right) \bm{u}_{\mathrm{sct}},
\end{align}
where $\theta$ and $\phi$ are the zenith and azimuth angles, respectively, in the spherical coordinates. Therefore, $\bm{W}$ can be written as
\begin{align}
 \bm{W} = \frac{\partial \bm{\Phi}_{\mathrm{sct}}^{\mathsf{H}}}{\partial\theta} \frac{\partial \bm{\Phi}_{\mathrm{sct}}}{\partial\theta} + \frac{\partial \bm{\Phi}_{\mathrm{sct}}^{\mathsf{H}}}{\partial\phi} \frac{\partial \bm{\Phi}_{\mathrm{sct}}}{\partial\phi}.
\end{align}
Each element of $\bm{W}$ is analytically obtained by using 
\begin{align}
& \frac{\partial \varphi_{\mathrm{sct},\nu\mu}}{\partial \theta} = \\
& \ \ \ \
\begin{cases}
 \sqrt{4\pi} h_{\nu}(kr) \mu \cot \theta Y_{\nu,\mu}(\theta,\phi), & \text{if} \ \nu=\mu \\
 \sqrt{4\pi} h_{\nu}(kr) \Big[ \mu \cot \theta Y_{\nu,\mu}(\theta,\phi) \\ 
  \ \ \ \ + \sqrt{(\nu-\mu)(\nu+\mu+1)}\mathrm{e}^{-\mathrm{j\phi}}Y_{\nu,\mu+1}(\theta,\phi) \Big], & \text{otherwise}
\end{cases}
\end{align}
and 
\begin{align}
 \frac{\partial \varphi_{\mathrm{sct},\nu\mu}}{\partial \phi} = \sqrt{4\pi} \mathrm{j}\mu h_{\nu}(kr) Y_{\nu,\mu}(\theta,\phi).
\end{align}
By using this weighting matrix $\bm{W}$, high-order coefficients are suppressed to small values. Then, the estimation accuracy is not largely affected by the setting of the truncation order $N$. 

\section{Experiments}

We conducted numerical experiments in a 3D free field to evaluate the proposed method. For comparison, the method based on kernel ridge regression described in Sect.~\ref{sec:priorwork} is used. The proposed method and the method based on kernel ridge regression are denoted as Proposed and KRR, respectively. 

\begin{figure}[t]
  \centering
  \centerline{\includegraphics[width=0.65\columnwidth]{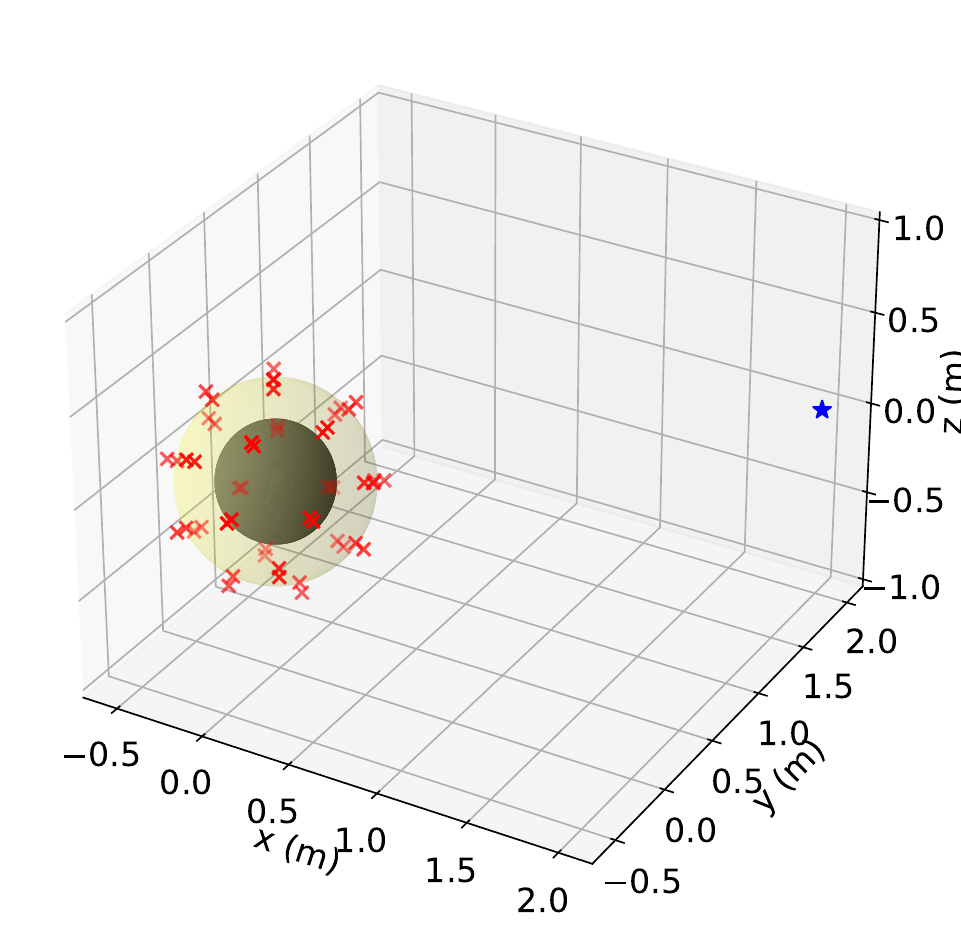}}
  \caption{Experimental setup. The spherical target region including a spherical scattering object was set. The red crosses and blue star indicate microphones and sound source, respectively.}
  \label{fig:exp_setting}
\end{figure}

As shown in Fig.~\ref{fig:exp_setting}, the target region $\Omega$ was a sphere of radius $R=0.5~\mathrm{m}$. An acoustically rigid spherical object of radius $0.3~\mathrm{m}$ was located inside $\Omega$. 25 omnidirectional microphones were distributed on two spherical surfaces of radius $0.5~\mathrm{m}$ and $0.55~\mathrm{m}$ by using spherical $t$-design~\cite{Chen:t-design}, which are indicated by red crosses in Fig.~\ref{fig:exp_setting}; therefore, the total number of microphones was 50. A point source (blue star) was at $(2.0, 2.0, 0.0)~\mathrm{m}$. Gaussian noise was added to the observed signals so that the signal-to-noise ratio becomes $40~\mathrm{dB}$. 

In Proposed, two truncation orders, $N=\lceil kR \rceil$ and $2\lceil kR \rceil$ with the radius $R$ of $\Omega$, were investigated with or without the weighting matrix $\bm{W}$, which is indicated as $\bm{W}$ and $\bm{I}$, respectively. The kernel function defined in \eqref{eq:kernel_diffuse} is used for both Proposed and KRR. The regularization parameters $\lambda$, $\lambda_1$, and $\lambda_2$ were chosen from $10^n$ with $n\in\mathbb{Z}([-15,9])$ based on the estimation accuracy. As an evaluation measure of the estimation accuracy, we define the following normalized mean square error (NMSE):
\begin{align}
 \mathrm{NMSE} := \frac{\int_{\Omega} |u_{\mathrm{inc}}(\bm{r}) - \hat{u}_{\mathrm{inc}}(\bm{r})|^2 \mathrm{d}\bm{r}}{\int_{\Omega} |u_{\mathrm{inc}}(\bm{r})|^2 \mathrm{d}\bm{r}},
\end{align}
where $\hat{u}_{\mathrm{inc}}$ denotes the estimated incident pressure distribution, and the integral is approximated as a summation at the evaluation points regularly distributed over $\Omega$ at intervals of $0.05~\mathrm{m}$. 

Fig.~\ref{fig:exp_nmse} shows the NMSE with respect to the frequency. The estimation accuracy of KRR significantly deteriorated owing to the effect of the scattering object because this method relies on the assumption that the target region is free space. In Proposed without the weighting matrix (Proposed ($\bm{I}$, $\lceil kR \rceil$) and Proposed ($\bm{I}$, $2\lceil kR \rceil$)), the NMSE was improved, but its performance was dependent on the truncation order. The difference of NMSE between Proposed ($\bm{I}$, $\lceil kR \rceil$) and Proposed ($\bm{I}$, $2\lceil kR \rceil$) was significantly large between $200$ and $600~\mathrm{Hz}$. The lowest NMSE was achieved by Proposed ($\bm{W}$, $\lceil kR \rceil$). Even when the truncation order was $2\lceil kR \rceil$, the deterioration of NMSE remained small. 

As an example, the estimated pressure and normalized error distributions of KRR and Proposed ($\bm{W}$, $\lceil kR \rceil$) on the $x$-$y$ plane at $z=0$ at the frequency of $300~\mathrm{Hz}$ are shown in Figs.~\ref{fig:exp_amp} and \ref{fig:exp_err}, respectively. High estimation accuracy was achieved over the target region $\Omega$ in Proposed, compared with KRR. 

\begin{figure}[t]
  \centering
  \centerline{\includegraphics[width=0.9\columnwidth]{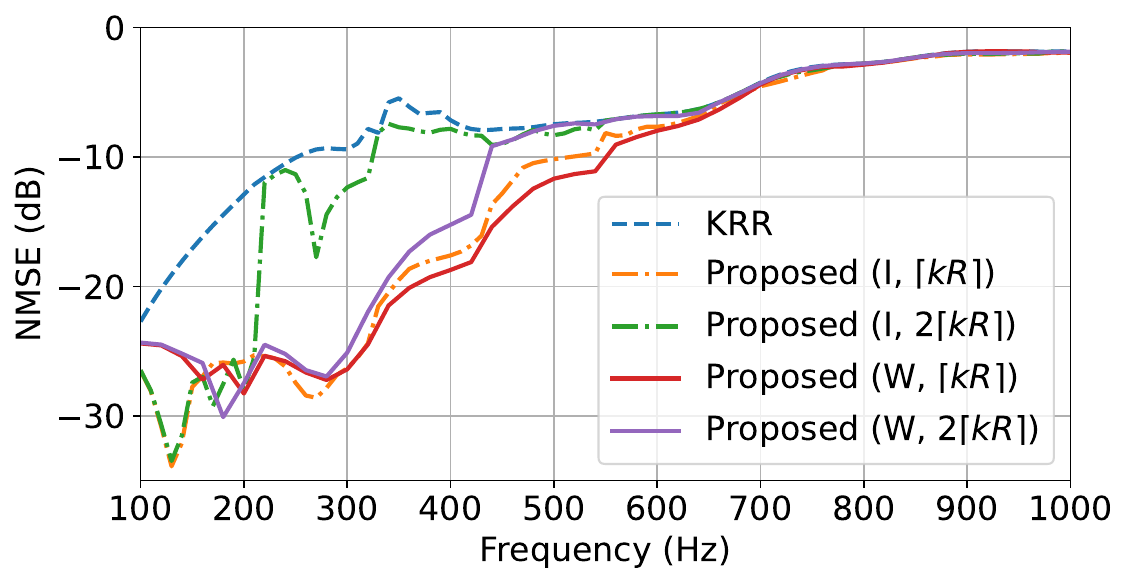}}
  \caption{NMSE with respect to frequency.}
  \label{fig:exp_nmse}
\end{figure}

\begin{figure}[t]
  \centering
  \subfloat[KRR]{\includegraphics[width=0.48\columnwidth]{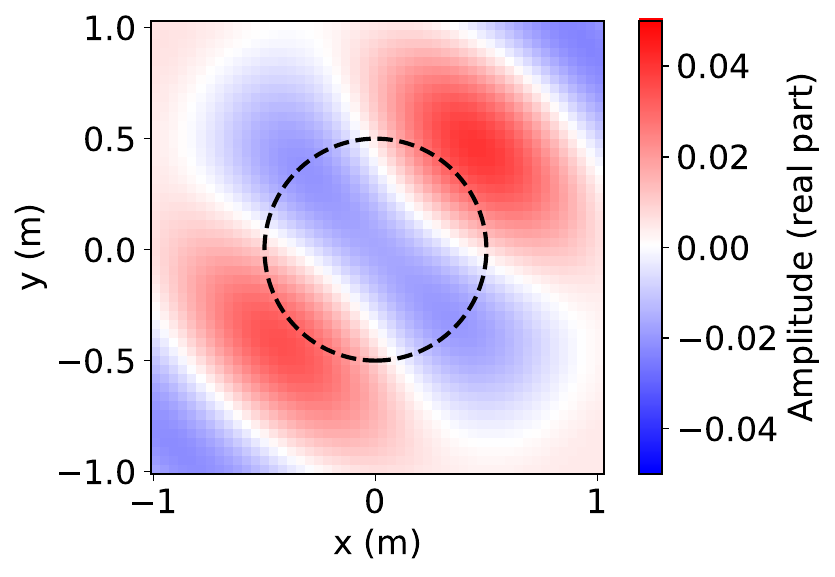}}%
  \subfloat[Proposed ($\bm{W}$, $\lceil kR \rceil$)]{\includegraphics[width=0.48\columnwidth]{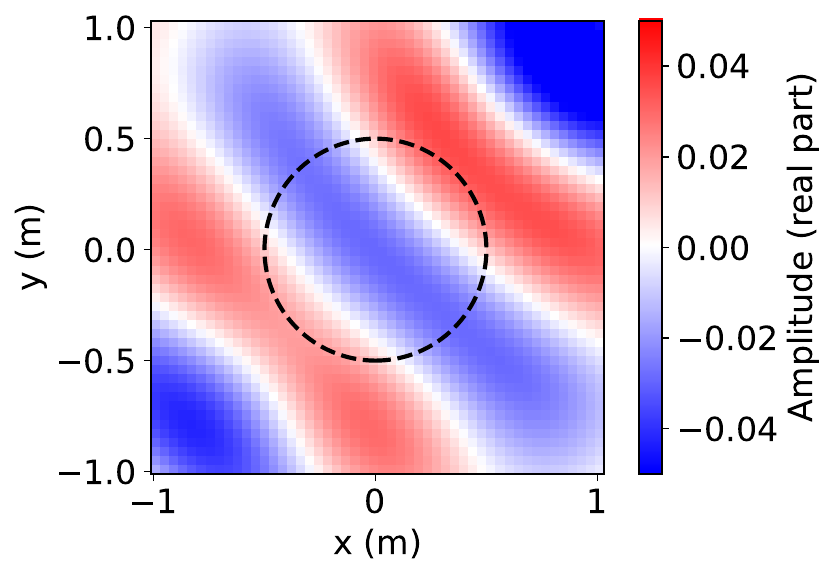}}
  \caption{Estimated pressure distributions at $300~\mathrm{Hz}$ on the $x$-$y$ plane at $z=0$. The dashed line indicates the target region.}
  \label{fig:exp_amp}
  \centering
  \subfloat[KRR]{\includegraphics[width=0.48\columnwidth]{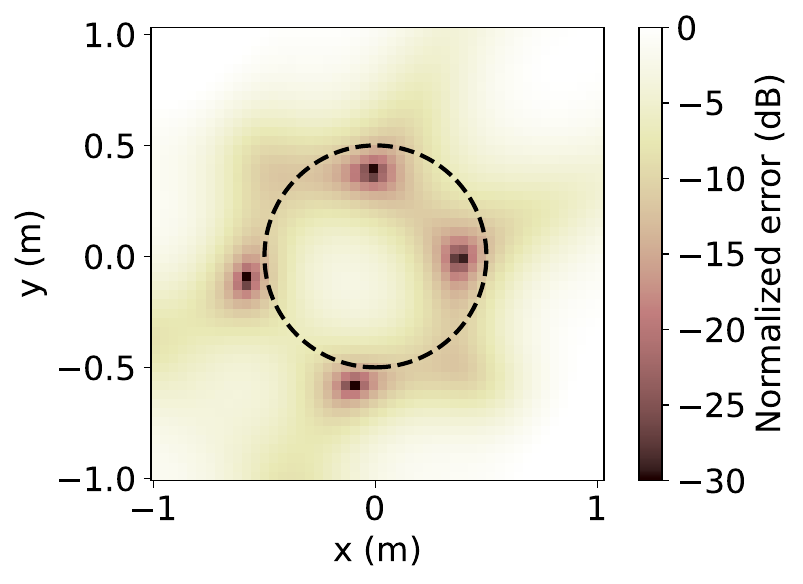}}%
  \subfloat[Proposed ($\bm{W}$, $\lceil kR \rceil$)]{\includegraphics[width=0.48\columnwidth]{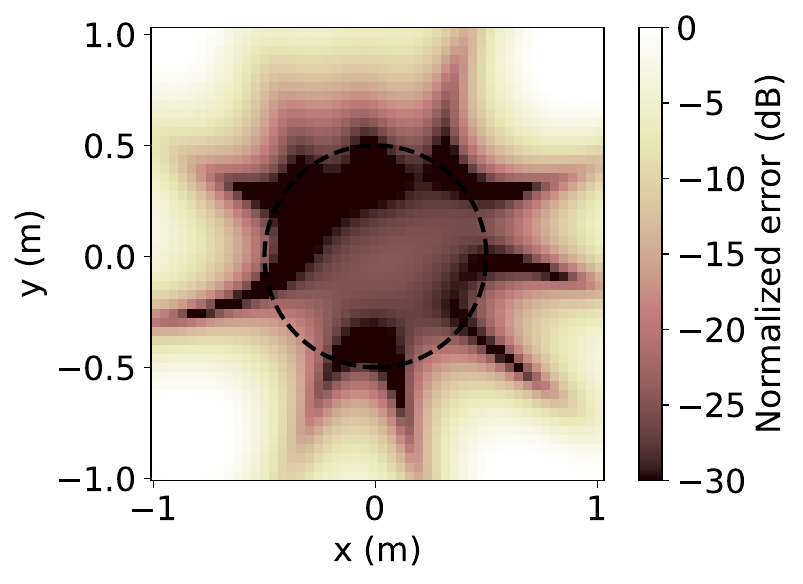}}
  \caption{Normalized error distributions at $300~\mathrm{Hz}$ on the $x$-$y$ plane at $z=0$. NMSEs of KRR and Proposed ($\bm{W}$, $\lceil kR \rceil$) were $-9.4$ and $-26.4~\mathrm{dB}$, respectively.}
  \label{fig:exp_err}
\end{figure}

\section{Conclusion}

We proposed a method to estimate the incident sound field in a region including scattering objects without precise knowledge or measurements of their properties. Our proposed method is based on the kernel ridge regression of the incident field with a separation from the scattering field represented by the finite-dimensional spherical wave function expansion. The optimization problem can be solved in a closed form; thus, the estimation can be performed by a convolution of a FIR filter. The weighting matrix for the expansion coefficients to induce smoothness of the scattering field in the angular direction alleviates the dependence of the estimation accuracy on the truncation order for the representation of the scattering field. In the numerical experiments, the proposed method achieved high estimation accuracy compared with the method based on kernel ridge regression without the separation. Future work will involve developing a method to determine regularization parameters in the estimator by using approximate knowledge of the scattering objects. 

\section{ACKNOWLEDGMENT}
\label{sec:ack}

This work was supported by JSPS KAKENHI Grant Number 22H03608 and JST FOREST Program Grant Number JPMJFR216M, Japan.


\bibliographystyle{IEEEtran}
\bibliography{str_def_abrv,koyama_en,refs}

\begin{thebibliography}{10}
\providecommand{\url}[1]{#1}
\def\UrlFont{\rmfamily}
\providecommand{\newblock}{\relax}
\providecommand{\bibinfo}[2]{#2}
\providecommand\BIBentrySTDinterwordspacing{\spaceskip=0pt\relax}
\providecommand\BIBentryALTinterwordstretchfactor{4}
\providecommand\BIBentryALTinterwordspacing{\spaceskip=\fontdimen2\font plus
\BIBentryALTinterwordstretchfactor\fontdimen3\font minus
  \fontdimen4\font\relax}
\providecommand\BIBforeignlanguage[2]{{%
\expandafter\ifx\csname l@#1\endcsname\relax
\typeout{** WARNING: IEEEtran.bst: No hyphenation pattern has been}%
\typeout{** loaded for the language `#1'. Using the pattern for}%
\typeout{** the default language instead.}%
\else
\language=\csname l@#1\endcsname
\fi
#2}}

\bibitem{Maynard:JASA1985}
J.~D. Maynard, E.~G. Williams, and Y.~Lee, ``Nearfield acoustic holography: I.
  {Theory} of generalized holography and the development of {NAH},'' \emph{J.
  Acoust. Soc. Amer.}, vol.~78, no.~4, pp. 1395--1413, 1985.

\bibitem{Bertin:CSAbook2015}
N.~Bertin, L.~Daudet, V.~Emiya, and R.~Gribonval, ``Compressive sensing in
  acoustic imaging,'' in \emph{Compressed Sensing and its Applications},
  H.~Boche, R.~Calderbank, G.~Kutyniok, and J.~Vybiral, Eds.\hskip 1em plus
  0.5em minus 0.4em\relax Springer, 2015.

\bibitem{Mignot:IEEE_J_ASLP2013}
R.~Mignot, L.~Daudet, and F.~Ollivier, ``Room reverberation reconstruction:
  Interpolation of the early part using compressed sensing,'' \emph{{IEEE}
  Trans. Audio, Speech, Lang. Process.}, vol.~21, no.~11, pp. 2301--2312, 2013.

\bibitem{Ribeiro:IEEE_ACM_J_ASLP2022}
J.~G.~C. Ribeiro, N.~Ueno, S.~Koyama, and H.~Saruwatari, ``Region-to-region
  kernel interpolation of acoustic transfer functions constrained by physical
  properties,'' \emph{{IEEE/ACM} Trans. Audio, Speech, Lang. Process.},
  vol.~30, pp. 2944--2954, 2022.

\bibitem{Park:JASA_J_2005}
M.~Park and B.~Rafaely, ``Sound-field analysis by plane-wave decomposition
  using spherical microphone array,'' \emph{J. Acoust. Soc. Amer.}, vol. 118,
  no.~5, pp. 3094--3103, 2005.

\bibitem{Teutsch:JASA2006}
H.~Teutsch and W.~Kellermann, ``Acoustic source detection and localization
  based on wavefield decomposition using circular microphone arrays,'' \emph{J.
  Acoust. Soc. Amer.}, vol. 120, no.~5, pp. 2724--2736, 2006.

\bibitem{Poletti:J_AES_2005}
M.~A. Poletti, ``Three-dimensional surround sound systems based on spherical
  harmonics,'' \emph{J. Audio Eng. Soc.}, vol.~53, no.~11, pp. 1004--1025, Nov.
  2005.

\bibitem{Iijima:JASA_J_2021}
N.~Iijima, S.~Koyama, and H.~Saruwatari, ``Binaural rendering from microphone
  array signals of arbitrary geometry,'' \emph{J. Acoust. Soc. Amer.}, vol.
  150, no.~4, pp. 2479--2491, 2021.

\bibitem{Koyama:JAES2023}
S.~Koyama, K.~Kimura, and N.~Ueno, ``Weighted pressure and mode matching for
  sound field reproduction: Theoretical and experimental comparisons,''
  \emph{J. Audio Eng. Soc.}, vol.~71, no.~4, pp. 173--185, 2023.

\bibitem{Zhang:IEEE_J_ASLP2018}
J.~{Zhang}, T.~D. {Abhayapala}, W.~{Zhang}, P.~N. {Samarasinghe}, and
  S.~{Jiang}, ``Active noise control over space: A wave domain approach,''
  \emph{{IEEE/ACM} Trans. Audio, Speech, Lang. Process.}, vol.~26, no.~4, pp.
  774--786, 2018.

\bibitem{Maeno:IEEE_J_ASLP2020}
Y.~{Maeno}, Y.~{Mitsufuji}, P.~N. {Samarasinghe}, N.~{Murata}, and T.~D.
  {Abhayapala}, ``Spherical-harmonic-domain feedforward active noise control
  using sparse decomposition of reference signals from distributed sensor
  arrays,'' \emph{{IEEE/ACM} Trans. Audio, Speech, Lang. Process.}, vol.~28,
  pp. 656--670, 2020.

\bibitem{Koyama:IEEE_ACM_J_ASLP2021}
S.~Koyama, J.~Brunnstr\"{o}m, H.~Ito, N.~Ueno, and H.~Saruwatari, ``Spatial
  active noise control based on kernel interpolation of sound field,''
  \emph{{IEEE/ACM} Trans. Audio, Speech, Lang. Process.}, vol.~29, pp.
  3052--3063, 2021.

\bibitem{Laborie:AESconv2003}
A.~Laborie, R.~Bruno, and S.~Montoya, ``A new comprehensive approach of
  surround sound recording,'' in \emph{Proc. 114th {AES} Conv.}, Amsterdam,
  Netherlands, 2003.

\bibitem{Samarasinghe:IEEE_ACM_J_ASLP2014}
P.~Samarasinghe, T.~Abhayapala, and M.~Poletti, ``Wavefield analysis over large
  areas using distributed higher order microphones,'' \emph{{IEEE/ACM} Trans.
  Audio, Speech, Lang. Process.}, vol.~22, no.~3, pp. 647--658, 2014.

\bibitem{Chardon:JASA2012}
G.~Chardon, L.~Daudet, A.~Peillot, F.~Ollivier, N.~Bertin, and R.~Gribonval,
  ``Near-field acoustic holography using sparsity and compressive sampling
  principles,'' \emph{J. Acoust. Soc. Amer.}, vol. 132, no.~3, pp. 1521--1534,
  2012.

\bibitem{Koyama:IEEE_J_JSTSP2019}
S.~Koyama and L.~Daudet, ``Sparse representation of a spatial sound field in a
  reverberant environment,'' \emph{{IEEE} J. Sel. Topics Signal Process.},
  vol.~13, no.~1, pp. 172--184, 2019.

\bibitem{Ueno:IEEE_SPL2018}
N.~Ueno, S.~Koyama, and H.~Saruwatari, ``Sound field recording using
  distributed microphones based on harmonic analysis of infinite order,''
  \emph{{IEEE} Signal Process. Lett.}, vol.~25, no.~1, pp. 135--139, 2018.

\bibitem{Ueno:IEEE_J_SP2021}
------, ``Directionally weighted wave field estimation exploiting prior
  information on source direction,'' \emph{{IEEE} Trans. Signal Process.},
  vol.~69, pp. 2383--2395, 2021.

\bibitem{Ueno:IWAENC2018}
------, ``Kernel ridge regression with constraint of helmholtz equation for
  sound field interpolation,'' in \emph{Proc. Int. Workshop Acoust. Signal
  Enhancement ({IWAENC})}, Tokyo, Sep. 2018, pp. 436--440.

\bibitem{Zotkin:ICASSP2017}
D.~N. Zotkin, N.~A. Gumerov, and R.~Duraiswami, ``Incident field recovery for
  an arbitrary-shaped scatterer,'' in \emph{Proc. {IEEE} Int. Conf. Acoust.,
  Speech, Signal Process. ({ICASSP})}, 2017, pp. 451--455.

\bibitem{Ahrens:IEEE_J_ASLP2022}
J.~Ahrens, H.~Helmholtz, D.~L. Alon, and S.~V.~A. Gar\'{\i}, ``Spherical
  harmonic decomposition of a sound field using microphones on a
  circumferential contour around a non-spherical baffle,'' \emph{{IEEE/ACM}
  Trans. Audio, Speech, Lang. Process.}, vol.~30, pp. 3110--3119, 2022.

\bibitem{Scholkopf:COLT2001}
B.~Sch\"{o}lkopf, R.~Herbrich, and A.~J. Smola, ``A generalized representer
  theorem,'' in \emph{Proc. Int. Conf. Comput. Learn. Theory (COLT)}, 2001, pp.
  416--426.

\bibitem{Horiuchi:WASPAA2021}
R.~Horiuchi, S.~Koyama, J.~G.~C. Ribeiro, N.~Ueno, and H.~Saruwatari, ``Kernel
  learning for sound field estimation with l1 and l2 regularizations,'' in
  \emph{Proc. {IEEE} Int. Workshop Appl. Signal Process. Audio Acoust.
  ({WASPAA})}, Oct. 2021, pp. 261--265.

\bibitem{Rasmussen:GPforML}
C.~E. Rasmussen and C.~K.~I. Williams, \emph{Gaussian Processes for Machine
  Learning}.\hskip 1em plus 0.5em minus 0.4em\relax The MIT Press, 2005.

\bibitem{Duraiswami:ICASSP2004}
R.~Duraiswami, D.~N. Zotkin, and N.~A. Gumerov, ``Interpolation and range
  extrapolation of hrtfs,'' in \emph{Proc. {IEEE} Int. Conf. Acoust., Speech,
  Signal Process. ({ICASSP})}, 2004, pp. IV--45--48.

\bibitem{Chen:t-design}
\BIBentryALTinterwordspacing
X.~Chen and R.~Womersley. (2004) Spherical $t$-design with $d=(t+1)^2$ points.
  Accessed: 2023-04-20. [Online]. Available:
  \url{https://www.polyu.edu.hk/ama/staff/xjchen/sphdesigns.html}
\BIBentrySTDinterwordspacing

\end{thebibliography}
%
%
%
%
%
%
%
%
%

\end{sloppy}
\end{document}